\documentclass[10pt,twocolumn,superscriptaddress,amsmath,amssymb,aps,pra]{revtex4-2}
\usepackage{amsmath}
\usepackage{graphicx}
\usepackage{booktabs} 
\usepackage{dcolumn}
\usepackage{physics}
\usepackage{comment}
\usepackage[T1]{fontenc}
\usepackage{mathtools}
\usepackage[mode = text]{siunitx}
\usepackage{newtx}
\usepackage{bm}

\newcommand\pitaya{Red Pitaya }
\newcommand\pitayas{Red Pitayas }
\newcommand\pitayaPeriod{Red Pitaya}
\newcommand\pitayasPeriod{Red Pitayas}
\newcommand\pyrpl{PyRPL }
\newcommand\pyrplPeriod{PyRPL}
\newcommand\beatnote{beat note }

\newcommand\uom[1]{\,\text{#1}} 

\usepackage[dvipsnames]{xcolor}
\definecolor{dark-blue}{rgb}{0,0.2,0.6}
\usepackage[colorlinks,%
            pdfusetitle,%
            urlcolor=dark-blue,%
            citecolor=dark-blue,%
            linkcolor=dark-blue]{hyperref}
\usepackage{orcidlink}

\usepackage{etoolbox}
\makeatletter
\pretocmd{\NAT@open}{\begingroup\color{\@citecolor}}{}{}
\apptocmd{\NAT@close}{\endgroup}{}{}
\makeatother

\begin{document}

\title{Long-term laser frequency stabilization with an FPGA-controlled scanning cavity}

\author{R.~Forti\orcidlink{0009-0009-9463-3397}}
\affiliation{Department of Physics, University of Trieste, 34127 Trieste, Italy}
\affiliation{Elettra Sincrotrone Trieste S.C.p.A., 34149 Trieste, Italy}

\author{R.~Panza\orcidlink{0009-0008-2059-4201}}
\affiliation{Department of Physics, University of Trieste, 34127 Trieste, Italy}
\affiliation{Istituto Nazionale di Ottica del Consiglio Nazionale delle Ricerche (CNR-INO), 34149 Trieste, Italy}

\author{A.~\surname{Muzi~Falconi}\orcidlink{0009-0008-5242-9016}}
\affiliation{Department of Physics, University of Trieste, 34127 Trieste, Italy}
\affiliation{Istituto Nazionale di Ottica del Consiglio Nazionale delle Ricerche (CNR-INO), 34149 Trieste, Italy}

\author{S.~Sbernardori\orcidlink{0009-0008-9184-8723}}
\affiliation{Department of Physics, University of Trieste, 34127 Trieste, Italy}
\affiliation{Istituto Nazionale di Ottica del Consiglio Nazionale delle Ricerche (CNR-INO), 34149 Trieste, Italy}

\author{A.~Vardè\orcidlink{0009-0008-2059-4201}}
\affiliation{Department of Physics, University of Trieste, 34127 Trieste, Italy}
\affiliation{Istituto Nazionale di Ottica del Consiglio Nazionale delle Ricerche (CNR-INO), 34149 Trieste, Italy}

\author{M.~Marinelli\orcidlink{0000-0002-1981-8182}}
\affiliation{Department of Physics, University of Trieste, 34127 Trieste, Italy}
\affiliation{Istituto Nazionale di Ottica del Consiglio Nazionale delle Ricerche (CNR-INO), 34149 Trieste, Italy}
\affiliation{Istituto Officina dei Materiali del Consiglio Nazionale delle Ricerche (CNR-IOM), 34149 Trieste, Italy}

\author{A.~Carini\orcidlink{0000-0003-4600-8123}}
\affiliation{Department of Engineering and Architecture, University of Trieste, 34127 Trieste, Italy}

\author{F.~Scazza\orcidlink{0000-0001-5527-1068}}
\email[E-mail address: ] {francesco.scazza@units.it}
\affiliation{Department of Physics, University of Trieste, 34127 Trieste, Italy}
\affiliation{Istituto Nazionale di Ottica del Consiglio Nazionale delle Ricerche (CNR-INO), 34149 Trieste, Italy}

\begin{abstract}
We present an FPGA‑based implementation of a scanning transfer cavity lock (STCL) for laser frequency stabilization, allowing for the simultaneous stabilization of multiple laser sources with respect to a single reference laser by means of a continuously scanned Fabry–Perot cavity.
By exploiting the FPGA architecture to simultaneously perform cavity scanning, peak detection, and feedback actuation, we minimize latency and allow independent control loops for several lasers within a single device, offering direct scalability. 
The system performance is analyzed through heterodyne measurements from short ($<1\,$s) to long ($\sim20\,$hours) timescales. 
The end-to-end locking performance is validated through atomic spectroscopy of ytterbium atoms in a magneto-optical trap, demonstrating sub‑MHz absolute frequency stability over several hours for the stability transfer across $\sim\!150\,$nm in the visible-wavelength range. 
Importantly, we demonstrate a novel fast-scanning approach based on acousto-optic modulator (AOM) frequency modulation, enabled by the low detection latency of our FPGA implementation. This increases significantly the effective locking bandwidth and reduces the intrinsic noise of the system with respect to standard piezo‑actuated scanning of the cavity length, allowing to reach sub-100\,kHz long-term stability and offering perspectives for laser-line narrowing.
Owing to its modularity, low cost and ease of implementation within the open-source \pyrpl firmware package for the STEMlab \pitaya platform, our architecture offers a compact and flexible alternative to existing STCL and locked-cavity implementations, providing a practical approach to the operation of a state-of-the-art cold-atom experiment without relying on any atomic references for laser stabilization.
\end{abstract}

\maketitle
\section{Introduction}
Laser frequency stabilization is an essential technique in atomic, molecular, and optical (AMO) physics~\cite{wieman_using_1991, nagourney_quantum_2010}. In many AMO experiments, it is required to simultaneously address broad and narrow optical transitions, each imposing distinct demands on the stability of the associated laser sources. Broad transitions, such as strong dipole‑allowed lines used for laser cooling and repumping, are typically compatible with the free-running MHz-scale spectral widths of common single-mode laser sources, requiring only long‑term drift compensation, often achieved by locking to atomic spectroscopic references~\cite{wieman_doppler_1976, hall_1981, bjorklund_frequency_1983,drever_laser_1983, black_introduction_2001, mccarron_modulation_2008} or precision wavemeters~\cite{couturier_laser_2018}. In contrast, the interrogation of narrow optical transitions necessitates laser linewidth reduction from the MHz scale down to the few-kHz, or even sub‑Hz regime, which is usually accomplished by locking to narrow atomic resonances~\cite{guttridge_direct_2016, atkinson_hyperfine_2019} or to high‑finesse ultrastable optical cavities~\cite{drever_laser_1983, salomon_1988, black_introduction_2001, ludlow_compact_2007, alnis_subhertz_2008, milani_2017}. 
The frequency stability of a locked laser can then be transferred to other lasers, even over wavelength separations of several hundred nanometers, using either actively stabilized transfer cavities~\cite{bohlouli-zanjani_2006,tonyushkin_2007_phase,lam_digital_2010,biesheuvel_2013,leopold_2016,zeng_2018} or optical frequency-comb-based virtual-beat and transfer-oscillator schemes~\cite{hagemann_2013,nicolodi_spectral_2014,barbieri_spectral_2019,savio_2025}.

The need for heterogeneous frequency locking methods is especially prominent in new-generation cold-atom experiments. Considerable effort has been devoted to cooling and manipulating atomic species with rich level structures, such as alkaline‑earth-like atoms~\cite{katori_1999, kuwamoto_1999}, lanthanide atoms~\cite{berglund_narrowline_2008,frisch_narrowline_2012}, or even more complex atomic species~\cite{bell_1999,uhlenber_2000, sukachev_2010,yu_2022,eustice_magneto_2025}, molecules~\cite{barry_2014,tarbutt_laser_2018, vilas_magneto-optical_2022} and multi‑species mixtures~\cite{hara_quantum_2011, hansen_2013, ilzhofer_2018, neri_2020, ravensbergen_2020,DeMartino_2025}, where a large number of independent laser sources must be stabilized. As the number of required lasers grows and direct frequency references are unavailable at some of the target wavelengths, traditional approaches lead to increasingly complex and costly setups, with limited scalability and tuning flexibility.

\begin{figure}
    \centering
    \includegraphics[width=1\linewidth]{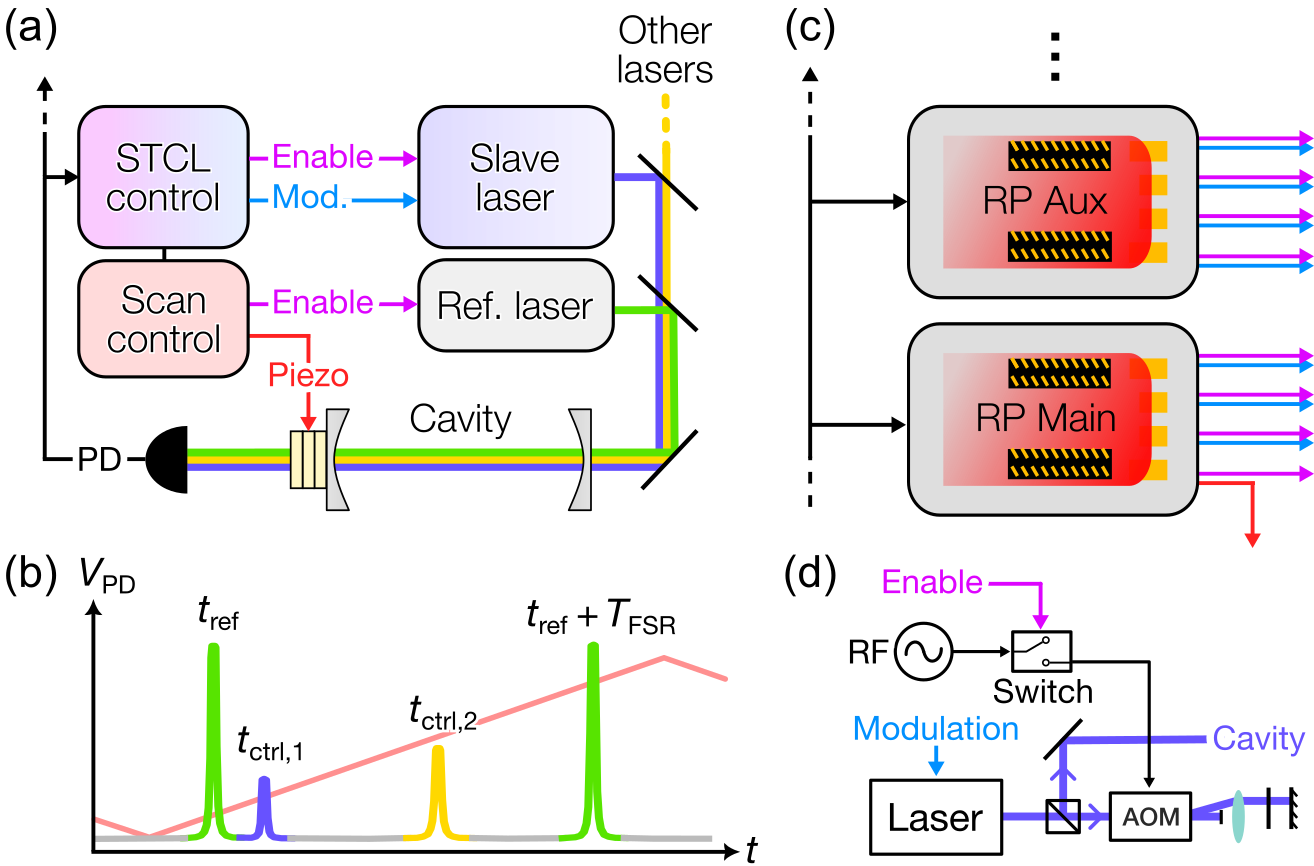}
    \caption{Simplified scheme of the STCL system. (a) Block diagram of our STCL system. The STCL control block actuates on the frequency of a slave laser, gating at the same time its injection into the optical cavity. The connected scan control block handles the scanning of the cavity length, and gates the injection of the reference laser into the cavity. A photodetector (PD) measures the cavity transmission signal and supplies it to the STCL control block. (b) Illustration of the typical cavity transmission signal, alongside the piezo displacement ramp (red), for multiple injected lasers: reference laser (green) and controlled lasers (violet, yellow). Each laser beam is coupled into the cavity only during the time window in which its resonance peak appears during the scan. The reference has two enabling windows to allow for the resonances of consecutive longitudinal modes (separated by one FSR) to be detected. (c) Connection of multiple \pitayasPeriod. Each device receives the photodetector signal and ramp trigger as inputs to implement up to four control blocks, supplying both the actuation and strobing signals. The main \pitaya also manages the cavity scan and the strobing of the reference laser. (d) Sketch of the optical setup for a slave laser: the beam goes through a double-pass AOM in cat-eye configuration, and is then directed towards the scanning cavity. The strobing is implemented by gating the RF signal that drives the AOM.}
    \label{fig:block scheme}
\end{figure}

A scanning transfer cavity lock (STCL)~\cite{zhao_1998_computer,burke_compact_2005,uetake_frequency_2009, seymour-smith_fast_2010,subhankar_microcontroller_2019,pultinevicius_scalable_2023} offers a conceptually simple and experimentally practical alternative for stabilizing several lasers against frequency fluctuations. 
By referencing all lasers to a single stable source through a continuously scanned Fabry–Perot cavity, the STCL enables the removal of frequency drifts across many lasers simultaneously. 
The general scheme of a STCL system is shown in Fig.~\ref{fig:block scheme}(a). The core of the system is an optical cavity with tunable length, continuously scanned by a piezo actuator.
Each laser frequency coupled into the optical cavity becomes resonant at specific cavity lengths, producing a fringe signal such as the one shown in Fig.~\ref{fig:block scheme}(b). 
The error signal for this locking scheme is derived from the relative positions of the reference and slave resonances within the scan. The slave lasers are then frequency-stabilized via feedback to keep their resonance peaks at fixed positions from scan to scan. 
By coupling multiple lights into the same cavity, the STCL scheme allows to independently lock several lasers to the same reference, provided that their resonances do not overlap significantly.

Despite its high versatility, the practical performance of the STCL scheme has been limited by control architectures in which time-critical tasks are not fully integrated into a single real-time system. 
In particular, STCLs have been experimentally realized using  microcontrollers~\cite{subhankar_microcontroller_2019,seymour-smith_fast_2010, burke_compact_2005} or CPU-assisted platforms~\cite{pultinevicius_scalable_2023} to handle the cavity scanning, peak detection in the cavity transmission spectrum and multiple independent control loops. 
However, these devices require a non-negligible time for calculations, which reduces the control bandwidth to $\sim\!100\,$Hz at best~\cite{pultinevicius_scalable_2023, subhankar_microcontroller_2019}, especially if a device has to control multiple lasers at a time.
Moreover, all previous implementations of STCLs have relied on ms-scale piezo-actuated scans, limiting the achievable control bandwidth to hundreds of Hz, irrespective of the electronic latency.

In this work, we present an FPGA-controlled implementation of a STCL, based on the STEMlab \pitaya boards, in which the control electronics no longer set the dominant bandwidth limit, which is instead determined primarily by the actuator.
Exploiting the versatility of the FPGA architecture, the cavity scan, peak detection, and feedback actuation on the slave lasers can be executed in parallel, rendering the overall calculation delay negligible. 
To this end, we develop an FPGA firmware and Python graphical user interface (GUI), building upon the widely used open-source \pyrpl project~\cite{neuhaus_pyrpl_2017, neuhaus_python_2024}. The \pyrpl firmware implements a variety of useful modules, that support rapid customization and functional extension, while the Python GUI enables intuitive control of all implemented FPGA functionalities.

Further, we introduce several improvements over  the standard STCL scheme~\cite{pultinevicius_scalable_2023, subhankar_microcontroller_2019, burke_compact_2005}. We employ dedicated acousto-optic modulators (AOMs) to strobe the different laser beams such that, during the cavity length scan, each injected laser is enabled only while its frequency is close to resonance with the cavity, thereby removing any cross-talk between control loops acting on independent laser sources. 
We demonstrate day-scale sub-MHz stability transfer across more than 200\,THz in the visible wavelength range. 
Slow drifts arising from nonlinearities of the piezoelectric response and by environmental refractive index changes are actively compensated using the reference laser and ambient pressure monitoring.
Enabled by the low latency of our control system, we also implement a faster scanning method based on AOM-assisted frequency modulation of each slave laser, showing a clear improvement over the standard piezo scan method in terms of locking bandwidth and noise suppression.
The implemented solution is designed to be highly scalable and cost-effective, since the optical setup requires only a single optical cavity, one AOM for each controlled laser, and one reference laser stabilized by an independent absolute reference. In particular, each \pitaya device added to the system can control up to four lasers 
[see Fig.~\ref{fig:block scheme}(c)], and all devices are managed through a single interface.

The rest of the paper is structured as follows: Section~\ref{chapt:pysics} describes the  working principles of the locking system; Section~\ref{chapt:setup} introduces the physical setup and hardware specifications; Section~\ref{chapt:algorithm} describes the algorithms implemented in the FPGA and the usage of the GUI; Section~\ref{chapt:results} presents characterization and validation measurements; Section~\ref{chapt:aom} introduces the AOM-enabled scanning strategy and presents its performance; the final Section \ref{chapt:conclusions} summarizes the results and proposes further envisioned improvements.

\section{Locking-scheme concept}\label{chapt:pysics}

The STCL scheme allows to compensate for frequency drifts of many laser sources with the use of a single measurement apparatus. 
The central component of the scheme is a Fabry-Perot cavity, whose length $L$ is linearly scanned with velocity $v$ over a micrometer-scale range by a piezoelectric actuator, $L(t)=L_0+vt$.
An injected laser becomes resonant with the cavity at a specific time across the scan that depends on its wavelength $\lambda$ -- specifically when $L$ is tuned to a multiple of $\lambda/2$ for a linear cavity. The laser resonance can be readily detected as a transmission peak through the cavity, and drifts can be compensated by actuating on the laser frequency so that the resonance peak position remains stable over repeated scans.   
Such control can be extended to stabilize multiple lasers through the same cavity. As the wavelength of each laser is generally different, the resonance peaks appear at different times across the scan. If the peaks are sufficiently separated from one another, the scan can be divided into separate ranges, each including a single resonance peak. The control system can measure the position of each peak independently from the others, allowing for the parallel detection of multiple laser resonances within a single scan.
The width of each resonance within the scan is determined by the amplitude of the scan and by the cavity finesse at each wavelength. Higher finesses lead in principle to narrower resonances, clearly benefiting the precision in the position estimation from the acquired cavity-transmission signal~\cite{kay1993statistical} and making space for fitting several lasers with non-overlapping resonances in a single scan. However, an excessive finesse limits the cavity scanning speed, reducing signal strength and resulting in asymmetric resonance shapes with side lobes associated with interference effect in the cavity ring-down~\cite{lawrence_1999}. Moreover, it introduces technical difficulties in the cavity injection alignment. Therefore, a tradeoff should be found where a similar finesse is obtained for all desired wavelengths, that allows to produce well separated resonances without introducing unwanted side-effects.

While it is in principle possible to stabilize many lasers within a single scanning cavity, in practice it is quite challenging to setup multiple simultaneous locks: 
\begin{itemize}
    \item[(i)] For each laser beam, imperfect coupling to the cavity fundamental Gaussian mode generates higher-order resonances, which may be located in the resonance range of other lasers, interfering with their correct detection.
    \item[(ii)] The resonance peaks of unlocked lasers, wandering uncontrollably in the scan range, may disturb the detection and locking of other lasers. 
    \item[(iii)] Two or more laser resonances may happen to demand the same position in the scan to operate at the correct frequency, requiring to introduce extra frequency offsets upstream the cavity, or to find a cavity length where all laser peaks are suitably isolated.
\end{itemize}
We resolve issues (i) and (ii) by gating the injection of each laser in the cavity by an optical switch.
This allows to couple each laser to the cavity only within its own sampling window during a scan, and we ensure that, within that range, the transmission signal is only generated by the correct laser, removing any cross-talk between different locks. 
We implement such staggered acquisition of several lasers by strobing each laser beam through a double-pass AOM, which also allows to finely tune the relative peak positions (up to tens of $\uom{MHz}$), thereby avoiding any scan-range overlaps [issue (iii) above]. 
While AOMs are not a strict necessity in the STCL setup, they further allow to implement an alternative strategy to piezo scanning (see Section~\ref{chapt:aom}), dramatically increasing the scanning rate and locking bandwidth. 

\subsection{Sources of disturbances and drifts}

In the STCL scheme, the laser frequency is indirectly measured by monitoring the position of a resonance peak with respect to the scan range. 
Besides changes of the source frequency, the peak positions can also be influenced by other components in the locking system, leading to inaccurate feedback and consequent deterioration of the stability over extended time scales.

\begin{itemize}
    \item[(a)] A change in the mean cavity length (i.e.,~a drift of $L_0$), caused by environmental fluctuations (e.g.,~thermal expansion of the cavity spacer) or by drifts on the average piezo displacement, leads to an uncontrolled shift of all resonance peaks.
    \item[(b)] A change in the amplitude of the piezo scan (i.e., a drift of the scan velocity $v$), caused by changes of the piezo voltage response, results in an uncontrolled shift of each peak proportional to its distance from the center of the scan.
    
    \item[(c)] A change of the refractive index in the cavity medium, due to fluctuations in ambient pressure, temperature and/or humidity~\cite{owens_optical_1967}, results in a wavelength-dependent differential variation of the optical path length between the mirrors~\cite{bohlouli-zanjani_2006,uetake_frequency_2009,subhankar_microcontroller_2019}. 
\end{itemize}
In general, any of these changes produce uncontrolled drifts in the error signal, and thus an erroneous modulation of the slave lasers frequencies that does not reflect any real frequency drift. 
It is therefore necessary to actively compensate for all shifts and fluctuations of the locking system itself.

The scan-related drift sources [(a) and (b)] can be suppressed by locking the piezo-scan voltage offset and amplitude to the reference laser, and then referencing the resonance of each controlled laser to those of the reference laser rather than to its absolute position within the scan.
In particular, acting on the offset voltage of the piezo scan, a resonance of the reference laser can be kept at a constant position across the scan, thereby removing any drift of the average cavity length. 
Additionally, acting on the scan amplitude, the distance from two consecutive resonances of the reference laser, namely the free spectral range (FSR), can be kept constant.
This effectively removes changes in the linear response of the piezo by fixing the displacement span of each scan.
These active controls cancel the long-term drifts arising from the piezo scan, yet they cannot counteract fast fluctuations of the cavity length. We remove any common-mode noise (i.e., noise that affects the position of all resonance peaks equally) by normalizing the slave laser peak position with respect to the reference laser peaks. Thus, instead of locking an absolute peak position with respect to the scan range, the control system locks the peak position with respect to two reference peaks separated by a FSR.

Changes in the refractive index of the medium between the cavity mirrors [(c)] can be prevented entirely by inserting the cavity in a controlled environment ~\cite{uetake_frequency_2009}, either under vacuum or a pressurized atmosphere. However, this solution increases the cost and complexity of the setup. An alternative solution is to monitor the pressure, temperature and humidity of the air in the cavity, and, after calibrating their effect on the positions of the cavity resonances, actively compensate for their drifts. 

\subsection{Aliasing of high-frequency noise}
The acquisition system converts the time-continuous signal of the laser frequency into a discrete signal. To avoid aliasing artifacts, a time-continuous signal should be low-pass filtered, removing all frequencies above half of the sampling frequency $f_s$ before sampling.
However, the signal acquired by our system (i.e., the position of the resonance peak) cannot be filtered before it is sampled. 
Thus, the STCL scheme is intrinsically bound to contain aliasing artifacts. Given a continuous-time signal $x(t)$ with Fourier transform $X(f)$, the control loop works with a sampled signal $x_{f_s}(t)=x(\lfloor tf_s\rfloor/f_s)$, which has the following Fourier transform 
\begin{equation*}
X_{f_s}(f)=\sum_{n=-\infty}^{\infty} X\!\left(f+n f_s\right)
\end{equation*}
Therefore, noise frequency components higher than $f_s/2$ are down-converted to low frequencies, 
which the control loop erroneously attempts to cancel. This produces an extra noise contribution at low frequencies, proportional to the high-frequency noise intensity and the magnitude of the control retroaction.
Even so, aliasing should not be a critical issue, since the spectral noise density of a laser typically decays as $1/f$, and the noise amplitude at drift frequencies for which the control loop is designed ($f \ll f_s$) should always be orders of magnitude larger than the noise resulting from aliasing of frequencies $f\simeq nf_s$. 
In addition, rough knowledge of the shape of the high-frequency noise spectrum of the laser is sufficient to model the expected aliasing distortion on the spectrum of the sampled signal~\cite{kirchner_aliasing_2005}, and identify the frequency for which aliasing becomes significant. This can thus be used as the limit bandwidth for the active control system, ensuring that the impact of aliasing remains negligible. 
In any case, working with a higher scan-rate $f_s$ helps reducing aliasing noise, 
and faster acquisition generally improves the drift removal, even if the bandwidth of the feedback loop is kept the same. 

An effective low-pass filtering of the original continuous-time signal is performed by the cavity due to the width of the resonance peaks. 
Noise on time scales much shorter than the cavity resonance width (expressed in scan time) only distorts the resonance shape, with minimal impact on the determination of the peak position from the acquired signal. Such filtering can be enhanced by choosing a cavity with lower finesse. However, a trade-off exists between the reduction of aliasing noise and the rejection of acquisition noise, the latter being enhanced by narrower cavity resonances. Additionally, broader resonances overlap more easily with one another, reducing the number of lasers that can be locked simultaneously.

\section{Experimental setup}\label{chapt:setup}
To assess the performance of our STCL control system, we realize a non-confocal optical cavity designed for broadband performance in the blue-green visible range, into which we couple three independent laser beams -- though the system can be readily extended to a larger number of lasers.
The cavity consists of two concave mirrors ($50\uom{mm}$ radius of curvature, $0.5$'' diameter) mounted in an Invar notched cylinder that serves as a stable spacer. It is designed to suppress vibration-induced cavity deformations and long-term drift of the mirror distance. 
The mirror coating (Thorlabs SA200-3B) provides a reflectance of 99.4(2)\% in the 350-560\uom{nm} range.
The mirrors are held in the Invar spacer by aluminum retaining-rings at a distance $L_0\simeq 82\uom{mm}$, resulting in a free spectral range FSR $\simeq 1.8\uom{GHz}$. One of the two mirrors is attached to a piezo ring actuator (PiezoDrive SR120610, capacitance $2.4\,\mu$F) and pressed against a rubber O-ring, allowing mirror displacements of a few tens of micrometers. The cavity finesse measured at 399\uom{mm} and 556\uom{mm} is $\,\mathcal{F} \approx 600$. The cavity is neither temperature stabilized nor shielded against acoustic noise or air pressure fluctuations, to accentuate environmentally induced drifts in the absolute frequency stability of the cavity.

As a reference laser source, we employ a $556\uom{nm}$ frequency-doubled VECSEL laser (Vexlum VALO SGH SF), externally stabilized to an ultra-low expansion (ULE) glass cavity (Menlo Systems ORC-Cylindric), characterized by a drift of less than 10\,kHz/day. A $399\uom{nm}$ diode laser (MOGlabs ILA 399), and another $556\uom{nm}$ fiber laser (Precilaser FL-SF-556-1.5-CW) are controlled by the STCL system. The slave laser frequencies are controlled by applying the analog outputs of the \pitayas to their frequency modulation inputs. The $556\uom{nm}$ slave laser is used to obtain a \beatnote signal with the reference laser, allowing to perform precise heterodyne measurements. The $399\uom{nm}$ slave laser is tuned to the $^1$S$_0 \rightarrow {}^1$P$_1$ transition of ${}^{174}$Yb. It is used to test the system for stability transfer between widely different wavelengths and to measure its absolute stability against drifts via atomic spectroscopy. 
Before coupling into the scanning cavity, each laser beam passes through an AOM (Gooch \& Housego 3080-125 and 3110-120) in a double-pass configuration, as shown in Fig.~\ref{fig:block scheme}(d). 
The RF signals that drive the AOMs are gated by digital outputs of the \pitayasPeriod, and they can be frequency modulated to allow for fine adjustments of the laser frequency reaching the cavity.
The three lasers are coupled into the cavity with the help of dichroic mirrors and polarizing beam splitters. 

A diagram of the control system connections is shown in Fig.~\ref{fig:block scheme}(a). The realized system would require a single \pitaya to manage the three lasers [see Fig.~\ref{fig:block scheme}(c)], but we employed two devices to test the modularity of the system. The main \pitaya controls the cavity scan and the $399\uom{nm}$ slave laser, while an auxiliary \pitaya controls the $556\uom{nm}$ slave laser. 
The first output of the  main \pitaya is connected to a piezo driver (PiezoDrive PDu150), which controls the piezo ring of the cavity, while the second output drives the frequency modulation input of the $399\uom{nm}$ laser. The signal sent to the cavity piezo is a symmetric triangular wave with a period of $4\uom{ms}$. The rising slope is used for the actual scan, while the falling slope is used to slowly bring the piezo back into position, avoiding any stress on the piezoelectric crystal due to sudden jumps in voltage. The auxiliary \pitaya is instead acting on the frequency modulation input of the $556\uom{nm}$ slave laser. Both \pitayas take as input the voltage from an amplified photodetector measuring the light transmitted through the cavity. 
We have modified the DAC output circuit of the \pitayas as described in \cite{pitaya_dac_performance} to reduce the noise of the analog outputs, helping to lower the noise floor of the control system. 

Since the control signals do not necessitate high bandwidth, the pulse-width modulation (PWM) pins of the \pitayas can be exploited to control the lasers, allowing each device to actuate on 4 lasers in parallel. The PWM pins generate an analog signal by repeatedly toggling a digital output and low-pass filtering it. The base FPGA firmware implements a PWM signal with an 8-bit precision and a bandwidth of $200\uom{kHz}$, limited by an RC circuit at the output. We modified the FPGA firmware to increase the resolution to 14-bit (the same resolution as the ADC and DAC modules), reducing the bandwidth to $10\uom{kHz}$. Yet, if the PWM pins are used, it is first necessary to buffer and filter their outputs, to remove any PWM noise at frequencies higher than $10\uom{kHz}$.

\begin{figure}[t]
    \centering
    \includegraphics[width=1\linewidth]{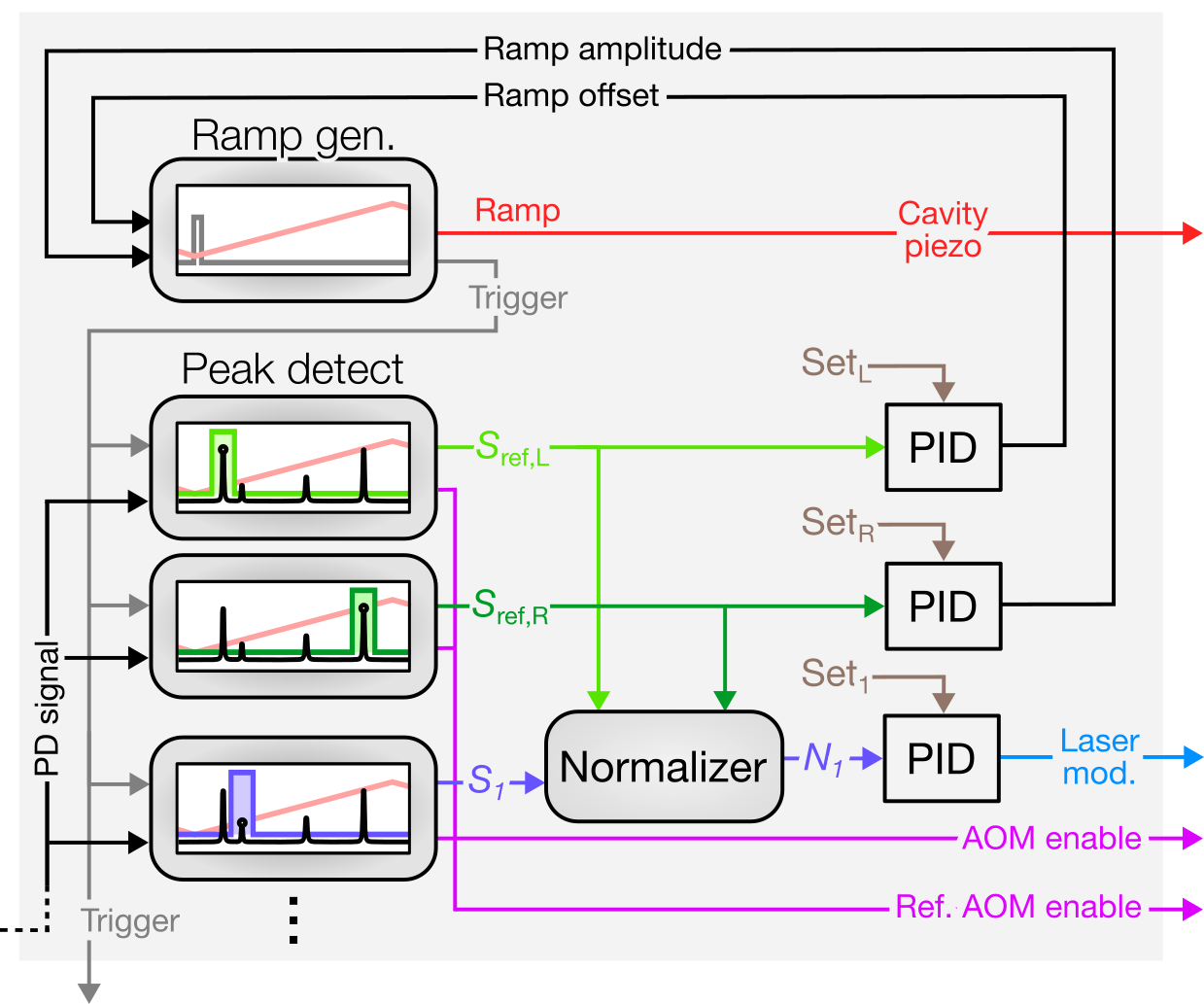}

    \caption{Block diagram of the internal operations executed by the main FPGA device. The \pitaya provides the voltage ramp to the cavity piezo and reads the PD signal of the cavity transmission. This signal is processed to identify the position of the relevant peaks, which are stabilized by digital PID modules actuating on the cavity ramp and on the slave laser frequency. The position of the slave laser is referenced to the normalized distance between the two (L and R) reference peaks. The \pitaya also outputs triggers to synchronize the cavity scan and to gate the laser lights. Auxiliary \pitayas implement a similar scheme without the ramp generator and without feedback on the reference peaks.
    }
    \label{fig:pitayaOperations}
\end{figure}

\section{Control Algorithm}\label{chapt:algorithm}
The software to execute the control system is based on \pyrplPeriod, a widely used software package that allows to control the various modules of the FPGA in a \pitaya device. Through its interface, the user can exploit the connected \pitaya for many purposes; among others, an oscilloscope, an arbitrary signal generator and multiple proportional-integral-derivative (PID) controllers. All of these modules are implemented inside the FPGA of each \pitayaPeriod, working in parallel. The entire \pyrpl project is open source, allowing to modify both the graphical user interface and the FPGA firmware. To implement the peak detection and scanning cavity protocol, we extensively modified the \pyrpl script and FPGA firmware. The resulting software is more application-specific, as some of the unused features had to be removed to free up the limited resources of the FPGA.

The operations executed by each \pitaya are shown in Fig.~\ref{fig:pitayaOperations}. The main device uses one of its arbitrary signal generator modules to generate the scan signal and a trigger signal, which is then sent to the oscilloscope module and to the other \pitayas to synchronize all other operations to the scan. The oscilloscope module acquires the signal from the cavity photodetector, and the resonance positions are obtained via the peak detection modules. The position of the slave peaks is normalized to the distance between the two reference peaks. In the main \pitayaPeriod, the reference peaks are also used to stabilize the offset and amplitude of the scan using PID modules. 
The normalized peak positions of the controlled lasers are instead sent to the corresponding laser frequency modulation inputs, after being processed by other PID modules. The peak detection modules also generate the gating signals that enable the relevant AOMs with correct timings during the scan.

\begin{figure*}[t]
    \centering
    \includegraphics[width=0.98\textwidth]{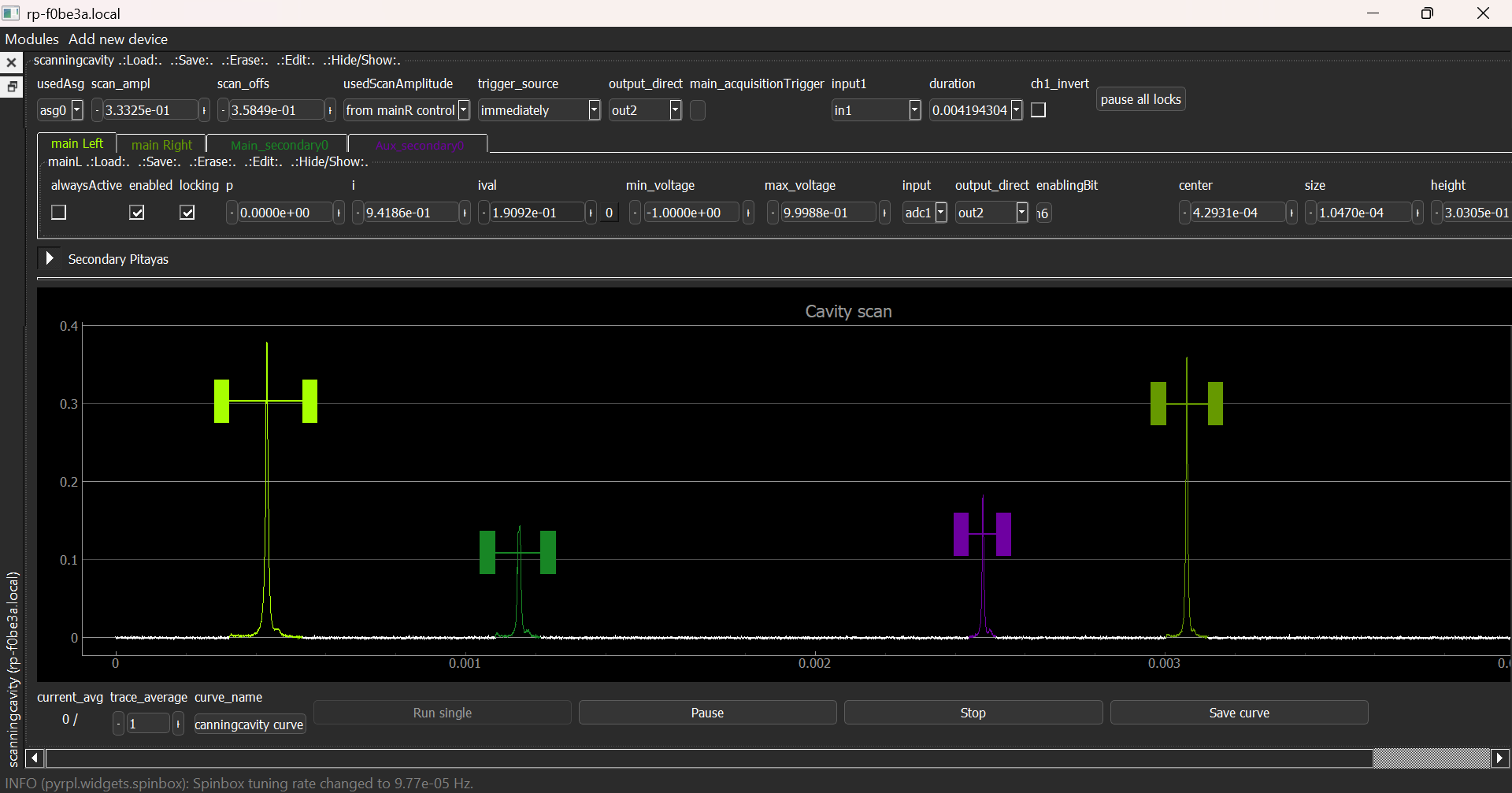}
    \caption{
    Screenshot of the graphical user interface of the newly implemented \texttt{ScanCavity} module in \pyrplPeriod. 
The interface allows to control all parameters of the STCL procedure and to visualize the cavity transmission during the scan.
Colored horizontal bars define the range during which each laser is coupled into the cavity: by coupling each beam only within its range, we avoid cross-talks between independent sources and detection of spurious peaks.
Each laser is locked to the center of its range, which can be resized and repositioned manually by drag-and-drop, even while the locks are engaged. Each slave laser PID setpoint can also be precisely adjusted through keyboard input.}
    \label{fig:ui}
\end{figure*}

\bigskip
\subsection{FPGA features}
Many of the operations required to implement the locking procedure are already supported by the existing FPGA modules, while others called for the implementation of custom modules -- in particular for peak detection and peak position normalization~\cite{footnote_FPGAspecs}. 
The peak detection module is implemented within the oscilloscope module, which handles the acquisition of signals. The module detects the position of the peak by finding the maximum value of the acquired signal within a specified time range of the acquisition. This calculation is repeated each time the oscilloscope receives an acquisition trigger, ensuring that the time range of the peak detector is always synchronized with the acquisition trigger. 
The peak detection algorithm includes a threshold parameter, such that peaks with amplitudes below this threshold are rejected as invalid. This becomes useful when the peak position is used as an input for a PID module. We implement a sample and hold, so that if a peak is not detected, the corresponding PID controller is halted, maintaining its last output. For normalized peaks, the control is also halted if any of the reference peaks is not valid. The sample\&hold feature can be exploited to temporarily halt the frequency locks during experiments, if required. For example, if the frequency of the reference laser needs to be momentarily changed, without affecting the controlled lasers. Also, if a laser has to jump between different frequencies, two separate peak detectors can be used. 

Since the positions of the controlled resonances have to be normalized with respect to the reference resonances, each auxiliary \pitaya needs to monitor the peak positions of the two reference resonances, alongside the resonance of its controlled lasers. Thus, six peak detection modules are implemented in each \pitayaPeriod.
The peak position $S_i$ of each slave laser is normalized by a dedicated module, which computes the normalized peak position
\begin{equation*}
N_i=\frac{S_i-S_\text{ref,L}}{S_\text{ref,R}-S_\text{ref,L}},
\end{equation*}
where $S_\text{ref,L}$ ($S_\text{ref,R}$) is the peak position of the left (right) reference resonance. The division operation is implemented in FPGA with the Vivado div\_gen IP core. 
The normalized peak positions are sent to PID modules, which use them as error signals to stabilize the laser frequencies. The setpoints of the PIDs are automatically set by the user interface to be at the center of each timing range. 

\subsection{User interface}
The \pyrpl user interface code has been significantly extended to implement the scanning and locking procedures, enable the control of multiple \pitayas within a single interface, and facilitate the development of additional modules and features. 
The \texttt{ScanCavity} module, shown in Fig.~\ref{fig:ui}, allows to control each parameter and property of the procedure, with an emphasis on ease of use in engaging each laser lock. The module is able to access parameters of other modules and to share information among multiple \pitayasPeriod.

Each laser can be enabled so that its resonance peak (or peaks) can be seen inside its range. Each range can be moved around the scan with drag-and-drop features, as well as with keyboard inputs. Moreover, the peak ranges can temporarily be allowed to overlap with one another, while each peak is still locked. 
These features allow for an easier setup of the locks and for easier movement and repositioning of the ranges, even after the lasers are already locked. 

\begin{figure*}
    \centering
    \includegraphics[width=0.75\textwidth]{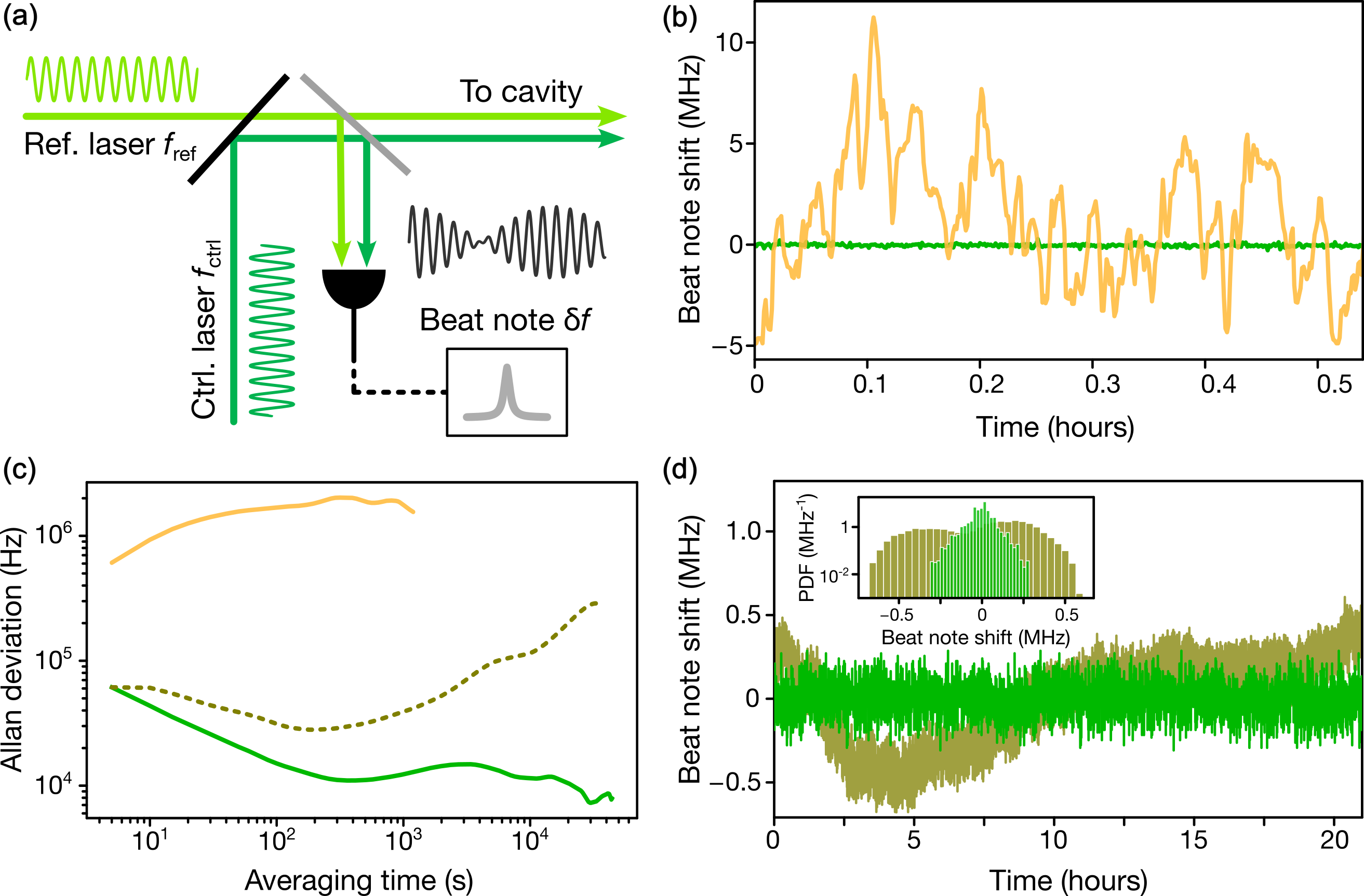}
    \caption{Heterodyne measurements between reference and controlled lasers at 556\,nm. (a) Sketch of the acquisition setup. The two laser beams are overlapped and interfered at a fast photodetector. (b) Comparison between \beatnote frequency deviations with the STCL control loop active (green) or inactive (orange). (c) Allan deviation of the acquired \beatnote frequency with the control loop active including the correction of scan drifts (solid green), excluding correction of the scan drifts (dashed olive) and inactive (solid orange). (d) Comparison between the closed-loop \beatnote stability over 20 hours with (green) and without (olive) active drift correction. The inset shows histograms of the beat note frequency deviations in the two cases.}
    \label{fig:\beatnote}
\end{figure*}

\subsection{Pressure drift compensation}
To compensate for refractive index drifts caused by environmental air pressure variations, we implement a simple feed-forward control scheme independent of the rest of the STCL system. A Bluetooth sensor (Ruuvi RuuviTag) is used to measure air pressure, temperature and humidity near the cavity, and a Python script reads the data from the sensor to modify the setpoint values of the \pitaya devices accordingly. 
This drift compensation is kept independent from the rest of the setup to allow for using different sensors and/or control algorithms.
From calibration measurements in our setup, the locked laser is found to drift proportionally to the air pressure changes, with negligible effects from typical temperature and humidity variations. The control algorithm continuously reads the pressure values sent by the sensor, and applies an offset and a gain to modify the setpoint of the peak position controller. The offset and gain values have been calibrated through the analysis of several-day-long acquisitions.

\section{Frequency stability performance}\label{chapt:results}

The locking performance is evaluated using two complementary methods: heterodyne \beatnote measurements between slave and reference lasers, and atomic spectroscopy. The former provide higher precision and access to high-frequency frequency fluctuations, whereas the latter precisely quantifies frequency drifts against an absolute atomic reference.

\subsection{Heterodyne measurements}\label{sec:heterodyne}

We monitor the \beatnote $\delta\!f(t)$ between the reference and the controlled $556\uom{nm}$ lasers. As displayed in Fig.~\ref{fig:\beatnote}(a), we use a beam sampler to deflect part of the light from both lasers towards a fast photodetector. 
The \beatnote frequency fluctuations is measured through a phasemeter (Liquid Instruments MokuLab), capable of performing fast tracking of a sinusoidal signal with a noisy frequency.
Without active frequency control, the \beatnote frequency is too unstable for the phasemeter tracking to operate for long intervals. Thus we measure the free-running \beatnote shift with an automated acquisition of the peak frequency of the \beatnote spectrum with a spectrum analyzer. This method affords a much lower sampling rate than phasemeter tracking, but it allows for much longer continuous acquisitions. 
The phasemeter acquisition is set to a sampling rate of $119\uom{Hz}$, while the spectrum analyzer acquisition is limited to $0.2\uom{Hz}$ (one acquisition every $5\uom{s}$).
To generate a beat note, the two lasers need to reach the photodetector simultaneously, hence the alternated-strobing feature has to be removed during the acquisitions, after making sure that no spurious cavity resonances can generate cross-talks between the two lasers.

Figures~\ref{fig:\beatnote}(b)-(d) display the results for long spectrum-analyzer acquisitions of the \beatnote deviation $\delta\!f(t)-\overline{\delta \!f}$, testing the performance of the control system against slow drifts. When enabled, the STCL system applies a PID control, with gain values set through the \texttt{ScanCavity} interface. For the measurements presented in Fig.~\ref{fig:\beatnote}, the time constant of the integrator is set to $5\uom{s}$, so that the entire bandwidth of the rejected noise falls within the sampling rate of the spectrum analyzer acquisition. In general, the system can be easily tuned to reject noise frequencies up to 100\,Hz. Considering that laser frequency noise decays typically as $\sim 1/f$ and that high-frequency fluctuations are filtered by the cavity response for $f \gtrsim 100$\uom{kHz}, the highest aliasing distortion seen by the control system within a 100\,Hz-bandwidth should be lower than $1\%$ of the input noise intensity (see also discussion on aliasing noise in Section~\ref{chapt:pysics}). 
Figure~\ref{fig:\beatnote}(c) displays the Allan deviation of the \beatnote frequency for free-running and locked slave laser. 
The locked-laser Allan deviation (solid green curve) displays a plateau for time scales longer than approximately 5 minutes, indicating the presence of flicker noise~\cite{allan_statistics_1966}. Importantly, since the Allan deviation does not begin rising again within the investigated timescale, we conclude that drifts are negligible for time scales up to $10\uom{}$ hours.
Additionally, measurements acquired with the active correction of cavity-scan drifts disengaged (see Section~\ref{chapt:setup}) are displayed in Figs.~\ref{fig:\beatnote}(c,d) (olive curves). The effect of cavity length and piezo-response drifts on the frequency stability become clearly visible in the Allan deviation at time scales on the order of a minute. 
The root mean square (RMS) value of the overall frequency noise is estimated as $\delta\!f_\text{RMS}\simeq67\uom{kHz}$ from spectrum-analyzer measurements (in the bandwidth between 0 and 0.2\uom{Hz}), and $\delta\!f_\text{RMS}\simeq142\uom{kHz}$ from phasemeter measurements (100\uom{Hz} bandwidth).

To analyze the noise floor of the control system, i.e.~the absolute lower limit for the in-loop noise of the controlled laser, we perform heterodyne measurements with a single laser source, employing two different diffraction orders emerging from a double-pass AOM. 
In this case, the \beatnote frequency mean value $\overline{\delta\!f}$ is fixed by the frequency of the RF drive to the AOM. 
By considering the undiffracted beam as the reference laser and the diffracted beam as the controlled laser, we actively control the frequency of the latter by modulating the RF signal of the AOM. Thereby, the (otherwise Hz-stable) \beatnote will acquire the intrinsic noise of the control system.

\begin{figure}
    \centering
    \includegraphics[width=1\linewidth]{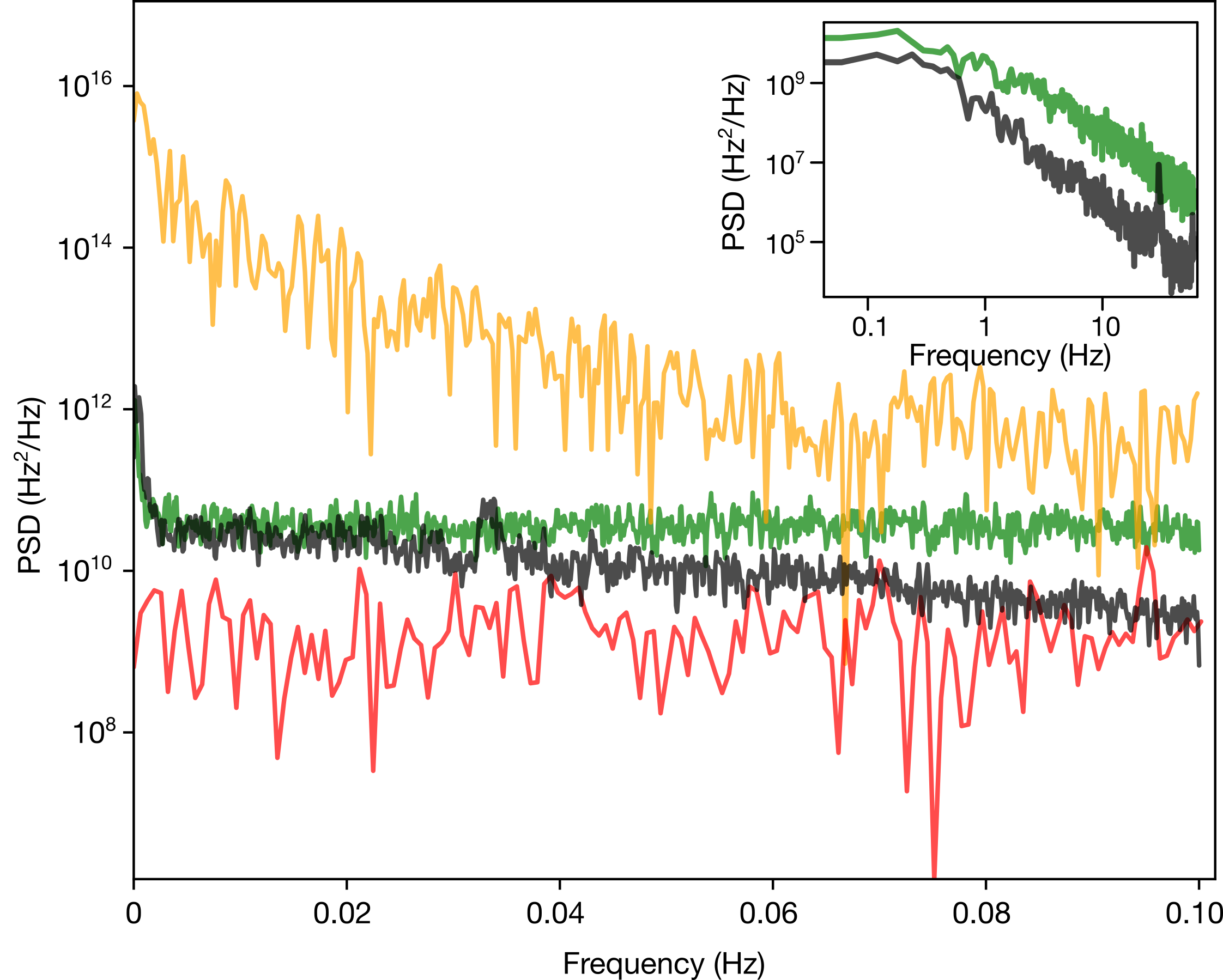}
    \caption{Power spectral density (PSD) of the \beatnote frequency between reference and controlled lasers at 556\,nm. Measurements for different settings of the STCL control are displayed: open loop (orange), closed loop with piezo-scanning (green), closed loop with AOM-scanning (red, see Section~\ref{chapt:aom}). The system noise floor in the piezo-scanning mode is also shown (black). Inset: a higher-frequency span of the PSD is shown, where the closed-loop piezo-scanning operation is compared to the noise floor across the $\simeq 100$\,Hz control bandwidth.}
    \label{fig:psd}
\end{figure}

Figure~\ref{fig:psd} summarizes the noise rejection performance of the control system, showing the power spectral density (PSD) of the \beatnote frequency deviations in the open-loop and closed-loop configurations, extracted from phasemeter acquisitions. 
With the control activated, the measured frequency noise PSD reaches the noise floor at the lowest frequencies, indicating that the noise rejection is limited by the intrinsic noise of the control system. 
We remark that for lasers characterized by larger intrinsic frequency drifts than the 556\,nm fiber-laser employed here as a slave laser, the noise rejection could be larger and become limited by the feedback bandwidth rather than by the noise floor.

\subsection{Atomic spectroscopy measurements}

\begin{figure*}
    \centering
    \includegraphics[width=0.75\textwidth]{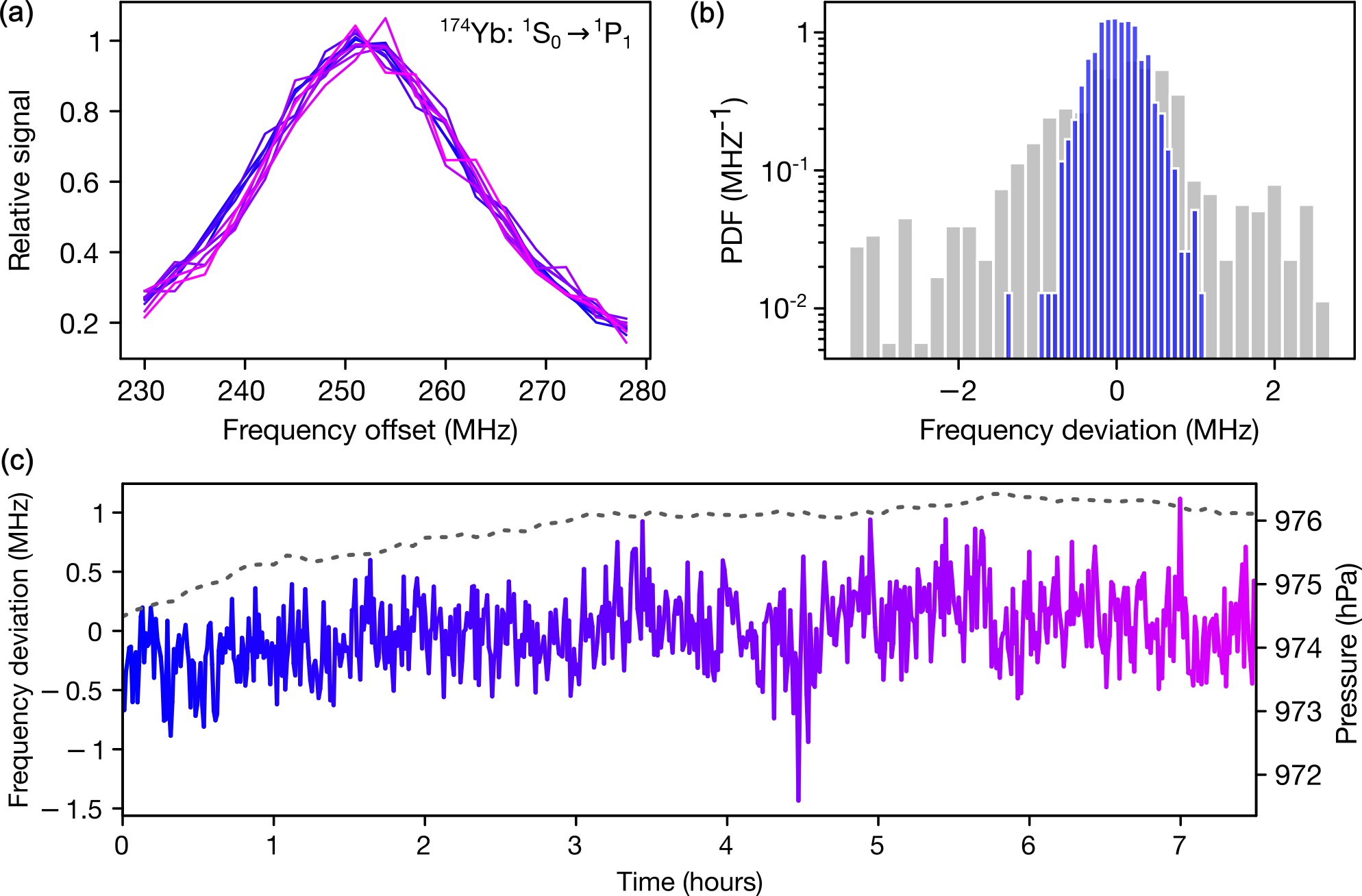}
    \caption{STCL long-term absolute frequency stability performance from atomic spectroscopy at 399\,nm. (a) Typical spectroscopic curves acquired over several hours of STCL operation by addressing the \mbox{$^1$S$_0 \rightarrow {}^1$P$_1$} transition of ${}^{174}$Yb with the slave 399\,nm laser. (b) Histogram of the laser frequency deviations extracted from Lorentzian fits to spectroscopic measurements over several hours, with active (blue) or inactive (gray) pressure compensation. (c) Frequency deviation of the 399\,nm laser recorded over time. The color coding references different spectroscopic curves shown in panel (a). The black dashed line displays the measured air pressure at the cavity location; a slight correlation with the frequency drift is still noticeable.}
    \label{fig:block spectroscopy}
\end{figure*}

We perform spectroscopic measurements of the \mbox{$^1$S$_0 \rightarrow {}^1$P$_1$} transition of ${}^{174}$Yb, using the controlled $399\uom{nm}$ laser as a probe beam to acquire absorption images of atoms held in a magneto-optical trap.
To estimate the laser frequency drifts, we execute repeated spectroscopic scans by tuning the probe beam frequency with an AOM, while the laser source frequency is actively controlled by the STCL system.  
Each spectroscopic scan takes approximately 40\,s and yields a curve for the detected atom number [see Fig.~\ref{fig:block spectroscopy}(a)]. 
By fitting each spectroscopic scan with a Lorentzian function, we can estimate the average laser frequency during that scan. Because of the relatively large linewidth of the transition $\gamma \simeq 29.1\uom{MHz}$, the measurement of the frequency is less precise than in heterodyne measurements. Yet, it enables us to quantify the absolute drifts of the controlled laser over long time intervals, in a case where the targeted stability transfer bridges a large wavelength difference $\sim 160\,$nm in the visible range.

Figures~\ref{fig:block spectroscopy}(b,c) display the spectroscopically extracted frequency deviation over time. The active compensation limits the effect of slow changes of the refractive index from air pressure variations, as can be seen in Fig.~\ref{fig:block spectroscopy}(b) where the frequency stability performances are compared for pressure compensation on or off. The laser frequency RMS deviation is kept below $1\uom{MHz}$ throughout $\sim\!10$ hours of continuous operation with no re-calibration, compared to a drift of several MHz in similar environmental conditions when pressure-compensation is deactivated. The pressure-change compensation procedure could be further improved by periodic automated re-calibrations to cancel sub-leading non-linear drifts, though the achieved level of stability is adequate for lasers addressing dipole-allowed, broad transitions such as the \mbox{$^1$S$_0 \rightarrow {}^1$P$_1$} line of ytterbium. 
Alternatively, placing the cavity in a controlled medium directly removes the sources of refractive-index changes, allowing to reach a performance similar to that demonstrated for a laser at a wavelength similar to the reference laser (see Section~\ref{chapt:results}\!.A).

\section{AOM-enabled fast scanning}\label{chapt:aom}

The double-pass AOMs used for gating the cavity injection of each laser beam can also be exploited to adjust its frequency. In this case, instead of scanning the cavity length, the frequency of the RF signals driving the AOMs can be scanned to measure the cavity resonance peaks. Since the typical AOM response to changes in frequency is on the order of the $\mu$s, the scan can be much faster than through piezo displacement of the cavity mirror, leading to a much higher sampling rate and servo bandwidth. Moreover, frequency-scanning of an AOM does not suffer from intrinsic drifts. 
However, the span that can be obtained by AOM scanning is on the order of tens of MHz, corresponding in our setup to about $3\%$ of a FSR. This means that the different laser resonances need to be close to one another in order to appear within the limited AOM-scan range, while the finesse has to be sufficiently high for all resonance peaks to remain well separated. 
Alternatively, one could set up the system so that only one laser is enabled in each AOM scan, albeit reducing the effective sampling frequency.

To arrange different laser resonances in close proximity, we deliberately excite higher-order resonances of the cavity (i.e., higher order transverse modes of each laser). While controlling the piezo displacement to scan the cavity length and adjust its mean value, one can act on beam alignments to find a condition in which every laser has a high-contrast resonance peak within a certain small range. For this strategy to be successful, the cavity must be far-from-confocal (and generally non-degenerate); additionally, increasing the power injected in the cavity could be necessary to enhance the visibility of high-order resonances. 

The frequency scan of the AOMs can then be implemented with a frequency-modulated signal generator for each AOM, the modulations being controlled by the second output of the main \pitaya (re-purposed from piezo scanning). The AOMs in our setup allow for a maximum frequency scan amplitude of $40\uom{MHz}$, over which their diffraction efficiency remains high. This is sufficient for fitting two resonances of the reference and slave $556\uom{nm}$ lasers, with a frequency difference of about $15\uom{MHz}$. 
A sawtooth waveform is used for the modulation control, since an abrupt frequency change does not affect the performance of the AOMs. 

\begin{figure}[t]
    \centering
    \includegraphics[width=1\linewidth]{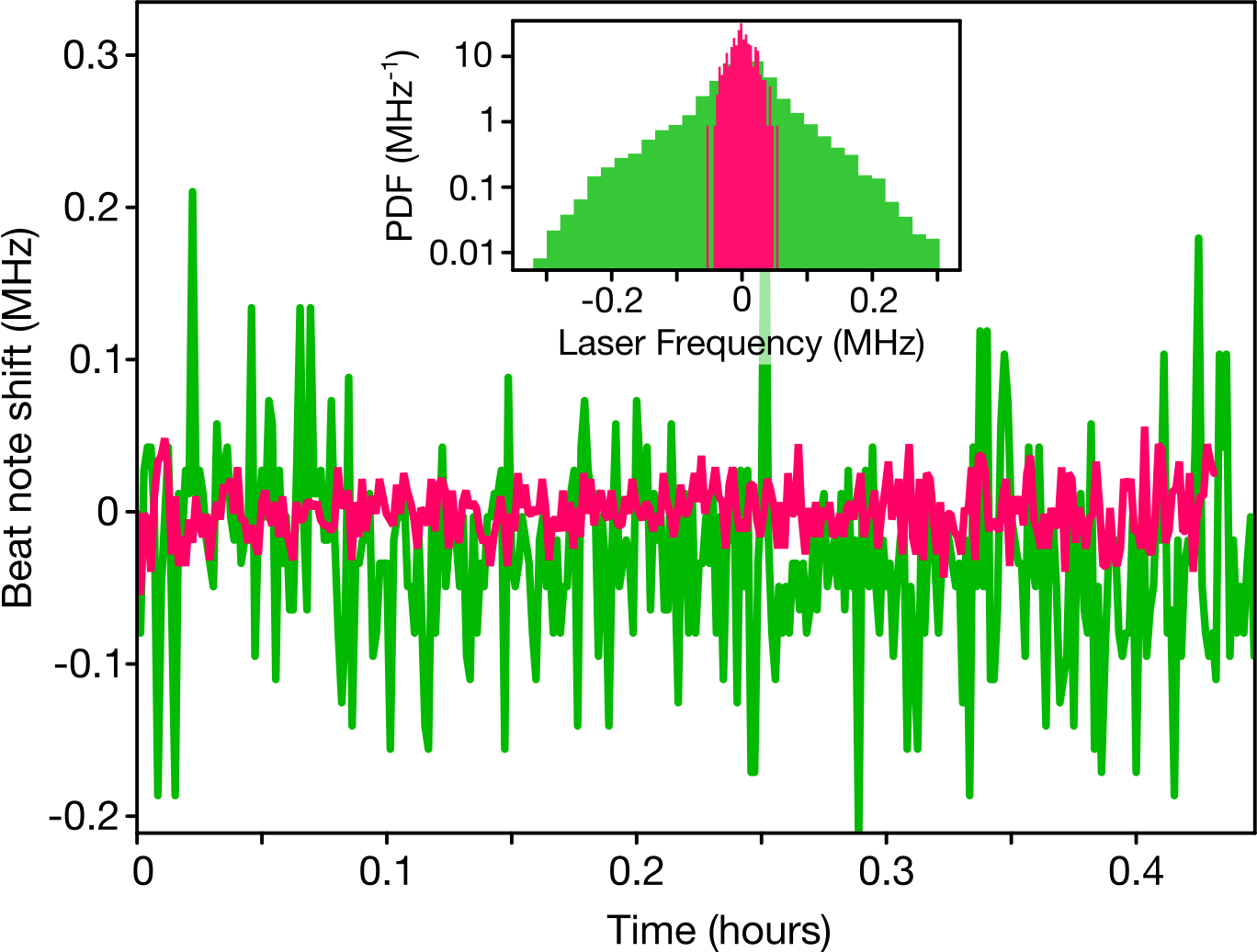}
    \caption{Heterodyne measurements between reference and controlled lasers at 556 nm in the STCL AOM-scanning mode (red) or piezo-scanning mode (green). The inset shows histograms of the \beatnote frequency deviations or the two modes, measured over an acquisition of $\sim0.5$ hours.}
    \label{fig:raw_AOM}
\end{figure}

Similarly to the piezo-scan mode of the STCL reported above, we tested the performance of the AOM-scanning mode by heterodyne measurements with two $556\uom{nm}$ lasers. 
Because atomic spectroscopy does not permit high-frequency measurements, and because drifts of the 399\,nm laser would in any case be dominated by refractive-index changes, we do not repeat spectroscopic measurements at 399\,nm with AOM scanning.
The frequency modulation of the two lasers is synchronous, thus the \beatnote is not affected by the continuous scanning. However, since the AOM diffraction efficiency depends on the RF drive, amplitude modulation generates extra tones in the \beatnote spectrum at the scanning frequency and its harmonics, that interfere with the phasemeter measurements. Therefore, only the spectrum-analyzer acquisition of the \beatnote peak frequency is employed here to monitor the lock stability.
In Fig.~\ref{fig:raw_AOM}, we show a typical measurement of the \beatnote frequency deviations obtained by setting the scanning frequency to $7.5\uom{kHz}$. We also show the corresponding \beatnote frequency noise PSD in Fig.~\ref{fig:psd}, comparing it with the results obtained in the piezo-scanning mode. Beyond enabling higher noise-rejection bandwidth, the AOM-scanning mode outperforms the noise floor of the piezo scan control throughout the entire measured frequency range.
This confirms the validity of the AOM-scanning strategy, which completely avoids all intrinsic disturbances associated to piezo scanning. The measured RMS frequency noise is $\delta\!f_\text{RMS}\simeq 19\uom{kHz}$ (0.2\,Hz bandwidth), demonstrating a more than three-fold improvement over piezo-scanning operation.

\section{Conclusions}\label{chapt:conclusions}
We introduced a scalable implementation of the STCL technique that takes advantage of the speed and parallel processing capabilities of FPGAs. The system reaches bandwidth of hundreds of $\uom{Hz}$, limited only by the speed of the scanning piezo, and provides sub-MHz frequency stability for uninterrupted operations of several hours, the remaining drift being dominated by environmental changes affecting the air refractive index differentially for distinct wavelengths. The proposed platform is cost effective, it is designed to simplify the locking procedures, and it is readily extensible to more complex locking schemes. Importantly, we showed that the STCL scheme can also be extended to the use of AOMs as the scanning devices, significantly increasing the feedback bandwidth of the system to several kHz and improving the noise rejection performance. Our system facilitates the operation of cold-atom experiments without relying on atomic references for laser frequency stabilization. In addition, our scheme permits to bridge isotope or hyperfine-state frequency gaps without requiring costly high-frequency acousto- or electro-optical modulators for applying the required laser frequency shifts.

The system is currently used within our cold-atom apparatus to stabilize a $399\uom{nm}$ source for laser slowing, cooling and absorption/fluorescence imaging on the \mbox{$^1$S$_0 \rightarrow {}^1$P$_1$} transition of ytterbium, where MHz-level frequency stability is required to suppress shot-to-shot fluctuations. Through atom fluorescence, we confirm that drifts are small enough to allow for several days of uninterrupted use before manual recalibration of the laser frequency is required. At present, the limiting factor for the long-term frequency stability transfer between lasers separated by a large gap > 200\,THz is the precise compensation of environmental pressure variations, which could be entirely resolved by placing the cavity in a pressure-controlled medium. 
Further developments may include the complete automatization of the locking procedure, alongside the implementation of particular frequency schedules. For instance, the AOM scanning technique may be used in combination with side-of-fringe locking to improve the feedback bandwidth, while still allowing for simultaneously locking multiple scanned lasers. 
Another extension of the system would concern the ability to perform controlled frequency jumps, whereby the frequency of a stabilized laser is suddenly shifted by a feed-forward step and then quickly re-stabilized at the target value by locking its resonance peak to a different setpoint in the scan range.
Finally, to broaden the applicability of the AOM-scanning technique, it would be beneficial to implement a longer optical cavity with smaller FSR in our system. Wideband acousto-optic deflectors, optimized for broad frequency operation rather than fast response, could enable scanning hundreds of $\uom{MHz}$ to cover an entire FSR. Alternatively, resonances of different lasers could be distributed over consecutive scans via the gating of cavity injections, enabling multiplexed control of different peaks that would otherwise overlap.

\bigskip 

\section*{Acknowledgments} 
We thank the Instrumentation and Detectors Laboratory at Elettra Sincrotrone for the constant support and guidance. We also thank Giacomo Cappellini, Stefano Finelli and Andreas Trenkwalder for useful discussions. This work has received financial support from the European Research Council (ERC) under the European Union’s Horizon 2020 research and innovation programme (project OrbiDynaMIQs, GA No.~949438) and from the Italian MUR under the FARE 2020 programme (project FastOrbit, Prot.~R20WNHFNKF). This work has also received funding from the European Union under the Horizon Europe programme HORIZON-CL4-2022-QUANTUM-02-SGA (project PASQuanS2.1, GA No.~101113690), the Digital Europe programme DIGITAL-2021-QCI-01 (project QUID, GA No.~101091408), and the Next Generation EU programme under the PNRR MUR project PE0000023-NQSTI and the MUR PRIN 2022 (project CoQuS, Prot.~2022ATM8FY). 

\section*{Data Availability} 
The modified \pyrpl project (containing the Python and Verilog codes for the user interface and the FPGA binary) is openly available in the GitLab repository \href{https://gitlab.com/arquslab-group/commontools/hardware/pyrpl}{https://gitlab.com/arquslab-group/commontools/hardware/pyrpl}. The data underlying the results presented in this paper may be obtained from the authors upon reasonable request.

\bibliography{cavity_scan}

\begin{thebibliography}{58}
\bibitem [{foo()}]{footnote_FPGAspecs}%
  \BibitemOpen
  \href@noop {} {}\bibinfo {note} {The current FPGA firmware uses $84\%$ of the
  available LookUp-Tables (LUT), $26\%$ of Flip-Flops, $47\%$ of BlockRAM
  (BRAM), and $14\%$ of Digital Signal Processing (DSP) blocks. Some of the
  default \pyrpl modules, such as the IQ and filter modules, were removed to
  free up space for the peak detection modules and one extra PID
  module.}\BibitemShut {Stop}%
\end{thebibliography}

\end{document}